# Cyclic Re-austenitization of Copper-bearing High-Strength Low-Alloy Steels Fabricated by Laser Powder Bed Fusion

Soumya Sridar, Yunhao Zhao, Wei Xiong*

*Physical Metallurgy and Materials Design Laboratory, Department of Mechanical Engineering and Materials Science, University of Pittsburgh, Pittsburgh, PA 15261, USA.*

* Corresponding author: *Tel: +1-(412) 383-8092, Tel: +1-(412) 624-4846, Email: w-xiong@outlook.com, weixiong@pitt.edu, URL: http://www.pitt.edu/~weixiong*

## Abstract:

For the first time, cyclic re-austenitization is carried out for additively manufactured high-strength low-alloy (HSLA) steels in order to refine the microstructure by reducing the prior austenite grain (PAG) size. In this work, HSLA-100 steels processed using laser powder bed fusion (LPBF) technique are subjected to several cycles of re-austenitization using quenching dilatometry. Microstructure characterization for every cycle revealed the presence of bainite, martensite and martensite/austenite (M/A) islands. From the analysis of the dilatometry curves and extensive microstructure characterization, it was found that till the $2^{nd}$ cycle of re-austenitization, both PAG size and martensite start ($M_s$) temperature get reduced, while the amount of bainite transformed decreased and the retained austenite content increased. Concomitantly, the highest microhardness along with peak nanohardness of the constituent phases was achieved at the $2^{nd}$ cycle. Conversely, from the $3^{rd}$ cycle, the microhardness, as well as the nanohardness of the constituent phases, are found to decrease due to an increase in the PAG size. This behavior is in contrast to the general tendency where a saturation limit is reached after the peak refinement is achieved. It is found that retained austenite can act as a pinning particle to obstruct the PAG boundary movement and its fraction is found to decrease from the $3^{rd}$ cycle. Hence, the increase in PAG size after the $3^{rd}$ cycle can be attributed to the destabilization of effective pinning particles to hinder the PAG boundary movement during the re-austenitization.

## 1. Introduction

In order to refine the martensite for enhanced strength with retained toughness of steels, the size of prior austenite grains (PAG) needs to be reduced, since this will result in a reduction of the block and packet size of the martensite [1]. In addition, this leads to an increase in the fraction of high angle grain boundaries in the as-quenched martensitic microstructure, that act as effective barriers for dislocation movement and hence, achieve strengthening [1–3]. This microstructure refinement can be accomplished using several methods which predominantly involves complicated and expensive deformation routes [4]. Ausforming is one such technique where the austenite is work-hardened, prior to the martensitic transformation to form fine martensite [5,6]. A high density of dislocations that forms during the deformation of the austenite, accelerates the nucleation of martensite accompanied by growth retardation [7]. However, if an external mechanical work is applied to refine the prior austenite grains, it will lead to a change in the shape of the component.

In order to circumvent this shortcoming, cyclic re-austenitization is a feasible option that can be applied to achieve grain refinement. It is a cost-effective and distortion-free technique to refine the PAG, without involving any deformation. Repeated heating and quenching for several cycles are performed either above $Ac_3$ (austenite finish) or between the $Ac_1$ (austenite start) and $Ac_3$ temperatures, depending on the material and application. This method was first applied by Grange [8] for medium carbon steels. Subsequently, it has been applied to a variety of steels, to achieve grain refinement in order to improve the strength and fatigue properties [4,7,9–17]. The refinement of martensite during cyclic re-austenitization occurs because of the successive transformation between austenite and martensite during the rapid heating and quenching [1]. In general, the new austenite grains nucleate at the high angle grain boundaries in the lath martensite, i.e., the block, packet, and prior austenite grain boundaries. However, as the austenite grain size decreases, the new austenite grains tend to form only at the PAG boundaries. The grain size saturates eventually, by attaining a balance between the refinement due to nucleation and the growth due to the completion of reversion [1]. An alternative theory was postulated for the grain refinement based on strain heterogeneity during cyclic re-austenitization of cold-worked austenitic stainless steels [18,19]. It was reported that a strain heterogeneity develops between the newly recrystallized and unrecrystallized regions leading to an irregular dispersion of the strain energy [18,19]. Hence, recrystallization is preferred against grain growth, thus, leading to grain refinement.

Additive manufacturing (AM) has attracted significant attention in recent years since it has several distinct advantages over conventional manufacturing processes such as high freedom to produce parts with complex geometry, reduced material wastage and high production flexibility [20,21]. The parts are built layer by layer using 3D numerical models during AM [22]. Laser powder bed fusion (LPBF) process is one of the commonly used AM method which involves melting successive layers of powder locally using a high-intensity laser beam [23]. Application of external mechanical work on AM builds to trigger recrystallization at elevated temperature cannot be a feasible option since AM is a near-net-shape process. Therefore, cyclic re-austenitization will be a promising approach to obtain a refined grain size after heat treatment in AM components for improved properties [24].

Owing to the difficulties in welding of steels with high carbon content for naval applications, copper-bearing high-strength low-alloy (HSLA) steels with low carbon content were developed [25]. These steels have gained interest as structural steels due to its high strength, excellent low-temperature toughness and good weldability. These properties are critical for applications such as construction of ship hull structures, offshore platforms as well as pressure vessel applications [25–28]. HSLA-100 (100 denotes the minimum obtainable tensile yield strength in ksi) steels belong to the class of copper-bearing HSLA steels that are cheap and involve low fabrication costs due to its superior weldability. This steel contains a higher amount of alloying elements such as Ni, Cu, Mn and Mo in comparison with the earlier grades that were in use for naval applications [26]. The high strength of HSLA-100 is derived from the hardening achieved due to the co-precipitation of BCC-Cu clusters in conjunction with $M_2C$ precipitates during aging [29]. The superior weldability results from the very low carbon content (< 0.06 wt.%) and hence, the composition is located in the Zone I of the Graville weldability diagram [30] leading to a reduced susceptibility for hydrogen-induced cracking. This leads to high strength and toughness in the heat-affected zone of the weldments [31].

The reduction in PAG size was achieved using cyclic re-austenitization for hot rolled and air-cooled HSLA-100 steel plates by Chapman *et al.* [16]. Moreover, an intrinsic cyclic heat treatment that was responsible for grain refinement and weak texture leading to a fine grained isotropic microstructure, during the processing of austenitic CrMnNi steel using electron beam powder bed fusion has been reported [32]. However, there are no reports available till now for cyclic re-austenitization of additively manufactured HSLA-100 steels. An evaluation of the feasibility of this method on AM steels will allow us to explore a new post-heat treatment strategy to enhance the quality of the AM builds. In this study, HSLA-100 steels processed using laser powder bed fusion (LPBF) are subjected to cyclic re-austenitization using quenching dilatometry and the effect of this treatment on the PAG size is investigated. With the aim of understanding the structure-property relationship, extensive microstructure characterization is performed, in order to correlate the refinement of PAG size due to cyclic-austenitization with the mechanical properties.

## 2. Experiments

Pre-alloyed HSLA-100 steel powders with composition (in wt.%) Al: 0.006, C: 0.046, Cr: 0.4, Cu: 1.44, Mn: 0.9, Mo: 0.8, Nb: 0.03, Ni: 3.47, Si: 0.19, were manufactured by Praxair Co., USA, using vacuum induction-melting atomizer with Ar gas, within a mesh size of -200 (74 μm) to -325 (44 μm). The d10, d50 and d90 particle sizes are 22, 32 and 47 μm, respectively. Cubes (1 cm³) of HSLA-100 steels were printed using EOS M 290 direct metal laser sintering machine with the factory default parameters for printing stainless steel 316L (Power=195 W; Scan speed=1083 mm/s; hatch spacing=0.9 mm; layer thickness=0.2 mm, size of build platform= 250x250 mm², temperature of build platform= 80°C). The X, Y and Z directions of the printed cube are the normal (scan), transverse (print) and build directions, respectively. The X-Z and X-Y planes are the build and transverse planes, respectively. The chemical composition of the as-built HSLA steel cube was measured using inductively coupled plasma optical emission spectroscopy (ICP-OES). The composition (in wt.%) was found to be Al: 0.006%, C: 0.042, Cr: 0.41, Cu: 1.32, Mn: 0.77, Mo: 0.84, Nb: 0.03, Ni: 3.45, Si: 0.19. The as-fabricated cubes were cut into cuboids of dimensions 5 x 5 x 10 mm³ using electric discharge machining (EDM, Mitsubishi MV2400S, Japan). The as-built HSLA steel consists of non-columnar martensite and bainite as the initial microstructure, with a PAG size of 7.5 μm and microhardness of 384±10 $HV_{0.1}$ [33]. Cyclic re-austenitization was performed for the cuboid samples using quenching dilatometry. The dilation as a function of temperature was measured using TA DIL805 A/D (TA Instruments Inc., USA) dilatometer. The sample was positioned between the two quartz pushrods that were connected to a linear voltage differential transducer (LVDT) to measure the dilation. An S-type thermocouple was spot welded at the center of the sample surface to measure the temperature. The samples were heated using an induction coil and quenched using Helium gas in order to achieve very high cooling rates (maximum of 2500 K/s). All the measurements were performed at a vacuum below $10^{-4}$ bar.

The equilibrium phase fraction as a function of temperature was calculated using the Thermo-Calc software [34] with the TCFE9 database for the pre-alloyed HSLA powder used for the LPBF process. Based on this thermodynamic calculation, the $Ac_3$ temperature was identified to be 754°C as shown in Fig. 1(a). Therefore, the austenitization temperature was chosen as 950°C, which is ~200°C above the calculated $Ac_3$ temperature to ensure complete austenitization. The time-temperature cycle used for cyclic re-austenitization is shown in Fig. 1(b). The samples were initially austenitized at 950°C for 40 mins followed by quenching (designated as $0^{th}$ cycle). Subsequently, the samples were re-austenitized by heating to the austenitization temperature and holding for a short duration of 1 minute followed by quenching, which corresponds to one cycle of re-austenitization. The re-austenitization cycle was performed repeatedly, which corresponds to the $1^{st}$, $2^{nd}$, $3^{rd}$ and $4^{th}$ cycles.

Extensive microstructure characterization for the AM HSLA-100 steel samples that were subjected to cyclic re-austenitization was performed in the build direction (Z direction of X-Z plane) using optical and electron microscopy. The samples were ground using 400 to 1200 grit SiC emery paper and polished with diamond and $Al_2O_3$ suspensions containing 1 and 0.01 μm particles, respectively. Subsequently, they were etched using 2% Nital (2 ml $HNO_3$ + 98 ml ethanol) for 10–30 seconds before observing the surface using a Zeiss Axio Lab A1 optical microscope. Further analysis of the phase evolution was performed using Zeiss Sigma 500 VP scanning electron microscope (SEM) with a field emission gun (FEG) source. Electron backscattered diffraction (EBSD) was carried out using FEI Scios Dual-Beam FIB-SEM with a step size of 0.08 μm. The automated reconstruction of the prior austenite grains from the martensite/bainite grains was accomplished using the ARPGE software package [35,36] with the EBSD data as input. The orientation relationship between the parent (austenite) and daughter (martensite/bainite) phases is used as the basis to determine the variants directly inherited by a single prior austenite grain. There are three orientation relationships possible during the martensitic transformation from austenite, that can be applied for the reconstruction of the PAG. They were proposed by Greninger and Troiano (G-T) [37], Kurdjomov and Sachs [38] and Nishiyama and Wassermann [39,40]. Using the G-T orientation relationship, the maximum reconstruction was obtained in comparison with the other orientation relationships and hence, it was used throughout this work. The size of the reconstructed PAG was measured using the linear intercept method [41].

The microhardness was measured using the Vickers microhardness tester (LECO LM800) with a load of 100 g and dwell time of 10 s. The reported values are an average of ten readings. The hardness of constituent phases was determined using nanoindentation (Hysitron TI900 TriboIndenter). The indentation was performed using a Berkovich tip with a half angle of 65°. A triangular load profile was used with a peak load of 5000 μN and a dwell time of 10 s. The load-displacement curve obtained after the indentation was analyzed using the Oliver-Pharr method [42] in order to determine the nanohardness. The surface topography of the indent was captured using an *in-situ* scanning probe microscope with a scan speed of 1 μm/sec.

## 3. Results and discussion

### 3.1. Quenching dilatometry

The dilation *vs*. temperature plots obtained after the quenching dilatometry measurements for different cycles of re-austenitization is shown in Fig. 2. The change in the linearity of the heating curve can be attributed to the transformation of the initial microstructure of AM HSLA steel to a complete austenitic structure. Similarly, the visible change in linearity in the cooling curve corresponds to the martensitic transformation. These deviations in the linearity of the heating and cooling curves can be attributed to the volumetric changes that occur during the phase transformation leading to dilation decrease of the sample. Besides, a minor change in linearity in the cooling curve before the start of the martensite formation corresponds to the start of the bainitic transformation. However, this change is not clearly noticeable to distinguish between the finish point of bainitic and start point of martensitic transformations. This minor deviation indicates that the bainitic transformation is partial and its fraction will be lower than that of the martensite in the final microstructure.

The amount of bainite formed at each cycle was calculated from the dilatometric curves using a geometrical method involving inflection point and tangents which is predominantly based on the commonly used Lever rule. The exact Lever rule could not be implemented due to the lack of clear distinction between the $B_f$ (bainite finish) and $M_s$ (martensite start) temperatures. As shown in Fig. 2(f), a perpendicular line is drawn at the inflection point (point B), the point where the second derivative of the change in length, $\Delta L$ with respect to temperature, T ($d^2\Delta L/dT^2$) equals to zero. The inflection point corresponds to the local maxima/minima in the first derivative of the change in length with respect to temperature ($d\Delta L/dT$). The infection point identified from the minima in the first derivative for each cycle of re-austenitization in the temperature range where martensitic transformation occurs are shown in Fig. 3. The perpendicular line intersects the tangents constructed for determining the $M_f$ (martensite finish) and $B_s$ (bainite start) temperatures (points A and C in Fig. 2(f), respectively). The ratio of the relative lengths is used to estimate the fraction of bainite (= length of AB / length of AC) and martensite (= length of BC / length of AC).

The critical points such as $Ac_1$, $Ac_3$, $B_s$, $M_s$ and $M_f$ temperatures were estimated from the dilatometric curves using the minimum deviation method [43]. In this method, a tangent is constructed from the linear region of the dilation *vs.* temperature curve and it is extrapolated into the non-linear region. The transformation temperature is identified as the point where the dilatometry curve deviates from linearity of the constructed tangent. The transformation temperatures determined using this method is shown in Fig. 4. It is evident that there is no major variation in the $Ac_1$, $Ac_3$, $B_s$ and $M_f$ temperatures determined for each cycle as shown in Figs. 4(a) and 4(b). Instead, a significant deviation is observed in the $M_s$ temperature calculated for each cycle (Fig. 4(c)) with the least temperature for the 2nd cycle of re-austenitization in comparison with the other cycles. The Ms temperature varies for each cycle depending on the prior austenite grain. Since each datapoint shown in Fig. 4(c) is calculated for all the cycles and not just the last cycle, the scatter of data is large. It is well known that the major factor that affects the Ms temperature is the chemical composition of the steel. In addition, it can be influenced by the PAG size significantly. Related studies [44,45] showed that the Ms temperature decreases as the PAG size decreases. The influence of PAG refinement on the $M_s$ temperature is found to be independent of the chemical composition [46]. The reduction in the $M_s$ temperature till 2 cycles followed by an increase in the subsequent cycles suggest that the maximum refinement of the PAG size occurs at the $2^{nd}$ cycle of re-austenitization.

There are several reasons available for this phenomenon, mostly explained based on the strengthening of austenite as the PAG gets refined [44]. Brofman *et al*. [47] proposed that the depression of $M_s$ temperature as the PAG size decreases can be explained based on the Hall-Petch strengthening of the austenite. It has also been reported based on several theories and experimental evidence that the dislocation density is inversely proportional to the austenite grain diameter [48]. Hence, the Hall-Petch strengthening arises from the increased dislocation density as the PAG size decreases, thus, strengthening the austenite. As the austenite grain size reduces, its resistance to plastic deformation locally and macroscopically increases. This, in turn, will directly inhibit the martensitic transformation by increasing the non-chemical free energy that opposes the transformation and hence, increases the $M_s$ temperature [49].

An alternative explanation was proposed by Olson et al. [50], based on the heterogeneous nucleation of martensite that requires suitable nucleating defects in the austenite. The defect might be a set of dislocations in the austenite/austenite interface [51] or the frozen-in vacancies formed during quenching from the austenitization temperature [52]. Hence, grain boundaries and other lattice imperfections act as nucleation sites that destabilize the austenite. A model was also developed based on probability which is given as $p = 1 - \exp(-\lambda v)$, where $p$ denotes the fraction of crystals containing martensite, $v$ is the grain volume and $\lambda$ is the probability for nucleation of martensite per unit volume, that depends on temperature [53]. From this equation, it is evident that the probability for nucleation decreases exponentially as the grain size decreases, which eventually reduces the $M_s$ temperature. Another plausible explanation was provided based on the dislocation density [54]. It was reported that the dislocation density in the austenite increases at temperatures above the $M_s$ temperature during the austenite-martensite reversion cycles. This leads to an increase of the shear stress required for the martensitic transformation and hence, reduce the $M_s$ temperature.

### 3.2. Microstructure characterization

The optical and SE-SEM (secondary electron) micrographs for each cycle of re-austenitization are shown in Fig. 5. The optical micrographs exhibit the presence of two different phases with varying contrast, one dark, and the other light. Similarly, the SEM micrographs show two different phase morphologies, one with laths and the other without visible laths. The former was designated as martensite and the latter as bainite from the varying contrast and morphology in the optical and SEM micrographs. Furthermore, the amount of bainite and martensite were calculated using image analysis by differentiating the phase contrast from the optical micrographs.

The inverse pole figure (IPF), image quality (IQ), and phase maps obtained using EBSD for each cycle are shown in Fig. 6. It is evident that the orientation of the grains is random from the IPF maps as well as the calculated intensities (Fig. 7) for all the cycles. The IQ map for each cycle is overlaid with the high angle grain boundaries ($15^o < \theta < 60^o$, red lines). Clusters of high angle grain boundaries (red islands in the IQ maps) have been observed after each cycle, which signifies the presence of martensite/austenite (M/A) islands. Though the difference in the dislocation densities of bainite and martensite is smaller, the presence of these two phases can be determined from the IQ maps. The brighter regions in the IQ map corresponds to the bainite with lower dislocation density and the darker regions signify the presence of martensite. The IQ curve for the BCC phase was found to be an asymmetric single large peak consisting of overlapping peaks for all the cycles as shown in Fig. 7. Wu *et al.* [55] has reported that the presence of an asymmetric IQ curve confirms the presence of two different microconstituents within the microstructure and hence, it can be fitted with 2 peaks (high-IQ and low-IQ peaks). The high-IQ and low-IQ peaks correspond to the bainite and martensite, respectively, and their amount were calculated from the area under the peak. From the phase maps, it is evident that the distribution of the retained austenite varies for each cycle. The retained austenite is present within the M/A islands as well as the grain boundaries. Since this study is mainly focused on the cyclic re-austenitization after an initial homogenization treatment at 950oC, Cu precipitates are not expected to form at that temperature.

The reconstructed PAG maps (Figs. 8 (a-e)) obtained for each cycle from the ARPGE software using the EBSD data as input show that more than 95% reconstruction has been achieved. The PAG size measured using the linear intercept method from the reconstructed PAG maps for each cycle (Fig. 8(f)) shows that the maximum refinement of the PAG was achieved at 2 cycles of re-austenitization. A decrease of around 25% from 0th to 1st cycle and nearly 36% reduction between the 1st and 2nd cycle was observed in the PAG size, which is a significant refinement obtained for AM HSLA-100. However, there is a 34% increase in the PAG size from the 2nd to the 3rd cycle of re-austenitization. This behavior is in contrast to the general tendency observed in other steels where a saturation limit is reached after the peak refinement is achieved [7,9].

Kernel average misorientation (KAM) as shown in Fig. 9 is a measure of local misorientations that can be retrieved directly from the EBSD data. The KAM represents the average misorientation around the measurement point with respect to a defined set of nearest neighbors or nearest and second-nearest neighbor points [56]. It also signifies the short-range orientation gradients within the grains and these local changes in the lattice orientation are considered as lattice curvature, which can be associated with the geometrically necessary dislocations (GND) [56]. GNDs are formed because of the stored dislocations associated with the non-uniform deformation. This creates a shear gradient that gives rise to a lattice rotation and a net Burgers vector for a set of dislocations. On the other hand, individual or group of dislocations with a net Burgers vector nearly equal to zero will not give rise to any significant lattice rotation. These are known as statistically stored dislocations (SSD) [57].

Figure 9 shows the KAM maps obtained from EBSD for each cycle of re-austenitization with misorientation up to 5o. The largest misorientation (red regions) appeared in regions where the FCC phase was indexed (mostly M/A islands). In general, blocks of relatively larger size exhibit almost zero lattice distortion in the core, and a misorientation between 1.5 and 2.5o is measured near the block boundaries [9]. It can be observed from Fig. 9 that the core regions showing nearly zero misorientation (blue regions) correspond to the areas with SSD and the contours around the core (green regions) with misorientation around 2.5°, corresponds to regions of high GND density stored in these locations. From the KAM map for each cycle, it is evident that the GND density is the highest for 2 cycles of re-austenitization. This demonstrates that as the PAG size reduces, there is an overall increase in the lattice distortion and thus, increases the GNDs. Figure 10(a) shows the average KAM plot for each cycle of re-austenitization. A predefined threshold value of 3° is set and points with values above the threshold are excluded since they are assumed to be belonging to adjacent grains or subgrains. It can be observed that the distribution of KAM is the widest for 2 cycles of re-austenitization and narrowest for $0^{th}$ cycle. The widening of the KAM distribution suggests that as the PAG size decreases, regions with high strain corresponding to the presence of GND increase.

The geometrically necessary dislocation density ($\rho_{GND}$) can be directly estimated from the KAM using the equation formulated by Kubin and Mortensen [58] which is given as follows.

$$\rho_{GND} = \frac{1.5\gamma\theta}{ub} \tag{1}$$

where $\theta$ is the misorientation angle that is normalized with the length between the points of the EBSD scan with the step size (u) and b is the Burgers vector. $\gamma$ is a constant that is dependent on the geometry of the block boundary. It takes a value of 2 or 4 for pure tilt or twist boundary, respectively. A value of 3 is used for a mixed type of boundaries in this work as suggested in Ref. [9]. Since the lattice curvature obtained from KAM is two-dimensional and ignores the lattice curvature along the surface normal, the $\rho_{GND}$ value can be underestimated. Hence, the whole expression is multiplied with a factor of 1.5, in order to compensate this error [9]. The $\rho_{GND}$ values calculated using Eq. (1) are shown in Fig. 10(b). The highest value of $\rho_{GND}$ is obtained for 2 cycles of re-austenitization. From these observations, it implies that as the PAG size is refined, the areas with high strain within a grain increase.

The amount of bainite and martensite calculated using the peak fitting procedure from the IQ curves for BCC as well as the amount of retained austenite obtained from the phase maps for every cycle is shown in Fig. 11(a). It is evident that the amount of bainite is the least and the fraction of retained austenite is the highest for the 2nd cycle of re-austenitization. The amount of bainite calculated from the dilatometry curves, optical micrographs, and EBSD (Fig. 11(b)) demonstrates that the trend in the variation with respect to the number of cycles is consistent, with the amount of bainite formed in the 2nd cycle being the least. However, the calculated amounts vary for the different methods used. The amount of bainite estimated using image analysis of the optical micrographs are the lowest possibly due to the smaller area of the sample being covered at higher magnifications when viewed under an optical microscope. It has been determined experimentally by Pavel [14], that the decrease in PAG size leads to a reduction in the amount of bainite formed in a given time as the bainite transformation kinetics gets altered. A longer time is required to transform to a given bainite fraction in smaller austenite grains, although the morphology of the bainite remains the same. Hence, this explains our finding that at 2 cycles of re-austenitization, the lowest amount of bainite has been transformed, which corresponds to the finest PAG size.

### 3.3. Hardness measurements

The microhardness as a function of re-austenitization cycles is shown in Fig. 12(a). It is evident that the peak hardness was achieved at 2 cycles of re-austenitization that can be correlated with the reduction of the $M_s$ temperature and PAG size.  As mentioned earlier, the PAG size gets refined after each cycle because of the nucleation of new austenite only at the PAG boundaries, as the austenite size decreases. It has also been claimed that after a certain limit, the grain size gets saturated by achieving a balance between refinement due to nucleation and grain growth because of austenitization [1]. The limit, i.e., the number of cycles, for refinement of PAG size depends on the material [14]. In this work, it has been found that the limit for refinement in PAG size is 2 cycles. However, in contrast to the postulate mentioned above, the refinement does not saturate and alternatively, the PAG size begins to increase, leading to a decrease in the microhardness in the subsequent (3rd and 4th) cycles.

Grange [8] has stated that effective refinement in PAG size can be ensured only if there is an efficient dispersion of particles that can inhibit the grain boundary movement. This gives rise to the question, what are the particles responsible for the refinement in the first 2 cycles of re-austenitization in AM HSLA-100 steels. From Fig. 11(a), it can be observed that the amount of retained austenite increases till 2 cycles of re-austenitization and further decreases in the subsequent cycles. It has been reported that as the PAG size decreases, there is an increase in the fraction of retained austenite which has an effect on the mechanical response of the material [59]. The growth of newly formed martensite is limited by the PAG boundary and hence, they possess high dislocation density. Therefore, an increase in the grain boundary density results in a higher fraction of retained austenite if the undercooling is not sufficient enough to promote the transformation [9]. The phase maps obtained from EBSD (Fig. 6) shows that besides the M/A constituents, retained austenite is mostly dispersed along the grain boundaries. Therefore, as the PAG size gets finer, the formation of the retained austenite increases until the $2^{nd}$ cycle of re-austenitization. The fraction of retained austenite decreases as the PAG size increases from the $3^{rd}$ cycle of re-austenitization and hence, there is a decrease in the microhardness. This proves that retained austenite is the effective pinning particles which hinder the grain boundary movement during cyclic re-austenitization of HSLA steels.

The nanohardness of the constituent phases as a function of the number of cycles of re-austenitization is shown in Fig. 12(b). Figures 12(c) and 12(d) present the scanning probe microscopic images of the nanoindents on the martensite and bainite, respectively. The difference in the nanohardness of the two phases (Fig. 12(b)) further confirms the presence martensite (with laths) and bainite (no laths) as seen from the optical and SEM micrographs (Fig. 4). It is evident from Fig. 12(b) that the peak nanohardness is achieved at the $2^{nd}$ cycle of re-austenitization for both the phases. Kenneth *et al.* [60] have reported that as the PAG size decreases, the dislocation density of the martensite increases. A similar phenomenon can be expected to occur in the bainite. The reduction in the size of the substructures of the martensite namely, blocks, packets and PAG is expected to increase the $\rho_{GND}$. The growth of the block is assumed to be the predominant factor since it controls the development of strain in the surrounding regions, thus controlling the $\rho_{GND}$ [9]. Moreover, as the PAG size decreases, the austenite gets strengthened due to the increase in the dislocation density [47]. Due to the strengthening, the martensite has to overcome higher shear stresses during its transformation, that necessitates the development of higher dislocation density to accommodate the plastic strain [9]. Therefore, the increase in the nanohardness of the martensite and the bainite can be attributed to the increase in the dislocation density as the PAG size gets refined. Chapman *et al*., [16] has reported that the grain refinement due to cyclic re-austenitization increases continuously as the number of cycles increase in hot rolled martensitic HSLA-100 steels. However, in additively manufactured HSLA-100 steels, we find that the grain refinement due to cyclic re-austenitization attains a peak at the $2^{nd}$ cycle of re-austenitization and subsequently decreases. This proves that as the processing conditions differ, the grain refinement behavior gets altered during cyclic re-austenitization.

In addition to this, the role of C distribution during the cyclic re-austenitization needs to be considered for the increase in the nanohardness of the bainite and martensite. Carbon diffusion occurs during quenching in low alloy steels with $M_s$ temperature above the room temperature [61]. It has been observed that carbides precipitate in coarse as-quenched martensite due to the tempering that occurs during the quenching, while it is free of carbides in fine lath martensite [62,63]. As the $M_s$ temperature is lowered due to the refinement of the PAG size, it can influence the distribution of C after the martensite formation. It has been reported that, coarse martensite blocks get tempered more and hence, carbon gets segregated in the form of plate-like defects (mostly carbides) [63], which can be correlated with the effect of PAG size on the dislocation density. It is known that as the PAG size gets finer, the dislocation density increases and hence, the C atoms that migrate from their interstitial positions during the tempering, get pinned by the dislocation. Therefore, they become less accessible to the formation of the carbides and increase the hardness of the martensite and bainite, which is evident from Fig. 12(b) for 2 cycles of re-austenitization.

Although cyclic re-austenitization is a conventional method to refine the prior austenite grain size, it has been applied to a near-net-shaped AM part successfully, in this work. However, there are several limitations to the application of this method. The first restriction is the type of material involved. The material should have a predominantly martensitic transformation upon rapid quenching. There are several reports where grain refinement has been achieved in martensitic steels using cyclic re-austenitization [9,12,15,17]. Certain titanium alloys also exhibit martensitic transformation, such as Ti-6Al-4V, which is a widely studied material for AM [64]. However, grain refinement through cyclic quenching is not successful for this alloy, due to the limited accumulation of dislocations and fast recovery and annihilation of those dislocations [24]. The second restriction is the thickness of the component. Small and thin components are more successful in achieving grain refinement since the samples get heated and quenched uniformly. However, for components with high thickness, it is very likely that the cooling rate may not reach the threshold value for martensitic transformation throughout the sample. Hence, it is difficult to achieve a uniform microstructure with cyclic re-austenitization for thick AM parts

## 4. Conclusions

In this work, HSLA-100 steels processed using LPBF technique is subjected to cyclic re-austenitization using quenching dilatometry, for the first time. The microstructure consisted of bainite, martensite, and M/A islands after each cycle. The peak microhardness was achieved at the $2^{nd}$ cycle of re-austenitization with the maximum refinement in PAG size. The PAG size reduced by nearly 50% from the $0^{th}$ cycle to the $2^{nd}$ cycle of re-austenitization. In addition to this, the nanohardness of the constituent phases such as martensite and bainite was the highest at the $2^{nd}$ cycle. As the PAG size gets refined, a reduction in the $M_s$ temperature was observed, along with a decrease as well as an increase in the amount of bainite and retained austenite, respectively. From the distribution of the retained austenite obtained from the microstructure characterization, it was deduced that it acts as particles that impede the coarsening of the PAG boundaries. The microhardness and the nanohardness of the constituent phases decreased from the $3^{rd}$ cycle of re-austenitization due to the coarsening of the PAG size and decrease in the fraction of retained austenite. Hence, it can be ascertained that the increase in PAG size from the $3^{rd}$ cycle is due to the destabilization of the effective pinning particle to hinder the grain boundary movement.

## Acknowledgments

The authors would like to gratefully acknowledge the financial support received from the Office of Naval Research (ONR) Additive Manufacturing Alloys for Naval Environments (AMANE) program (Contract No.: N00014-17-1-2586). The authors thank Miss. Yinxuan Li for her assistance in sample preparation through the Mascaro Centre for Sustainable Innovation (MCSI) summer internship program.

## Data Availability

The raw/processed data required to reproduce these findings cannot be shared at this time as the data also forms part of an ongoing study.

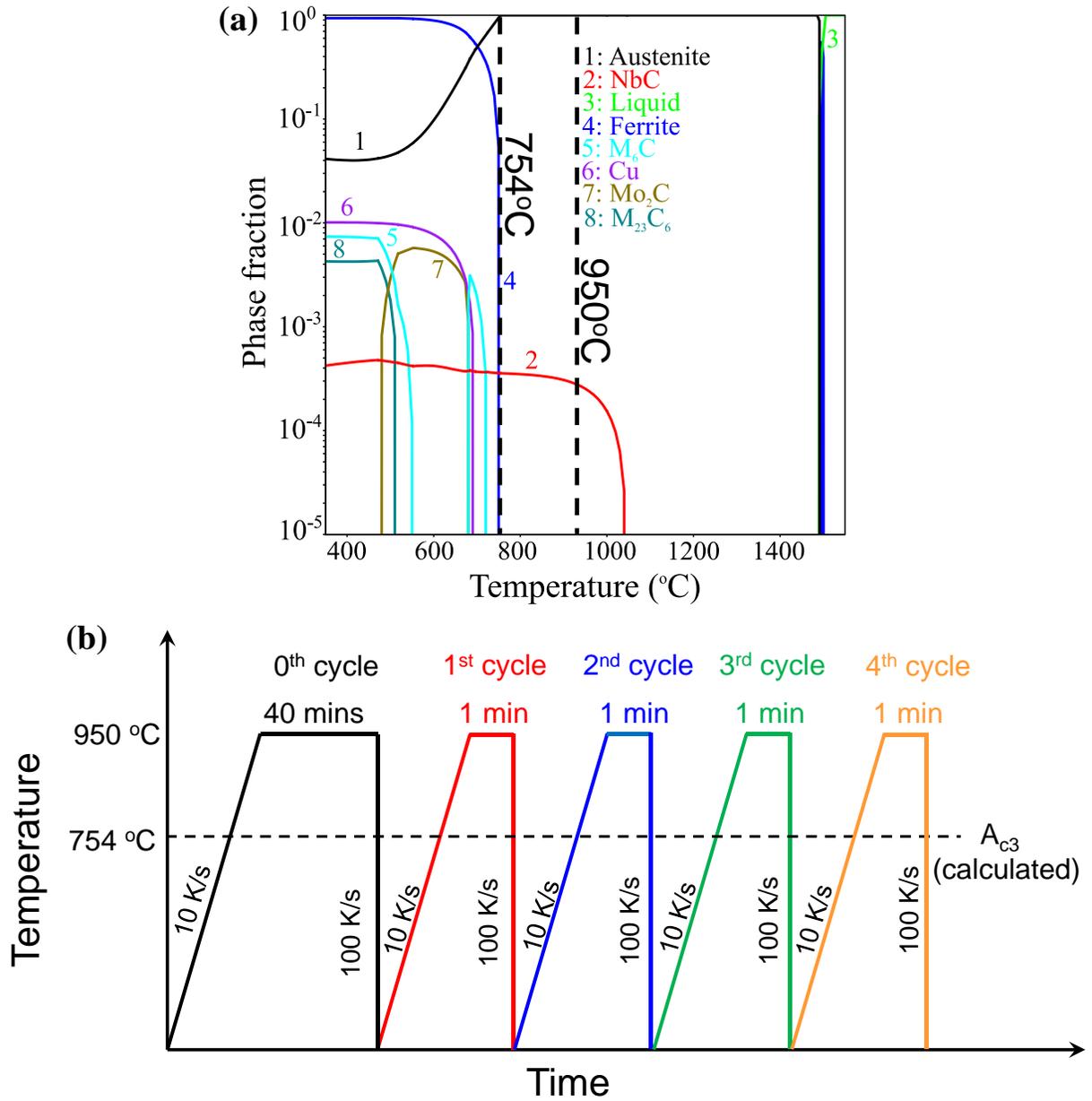

Fig. 1. (a) Calculated equilibrium phase fraction as a function of temperature for the pre-alloyed HSLA-100 powder and (b) time-temperature cycle used for cyclic re-austenitization using quenching dilatometry.

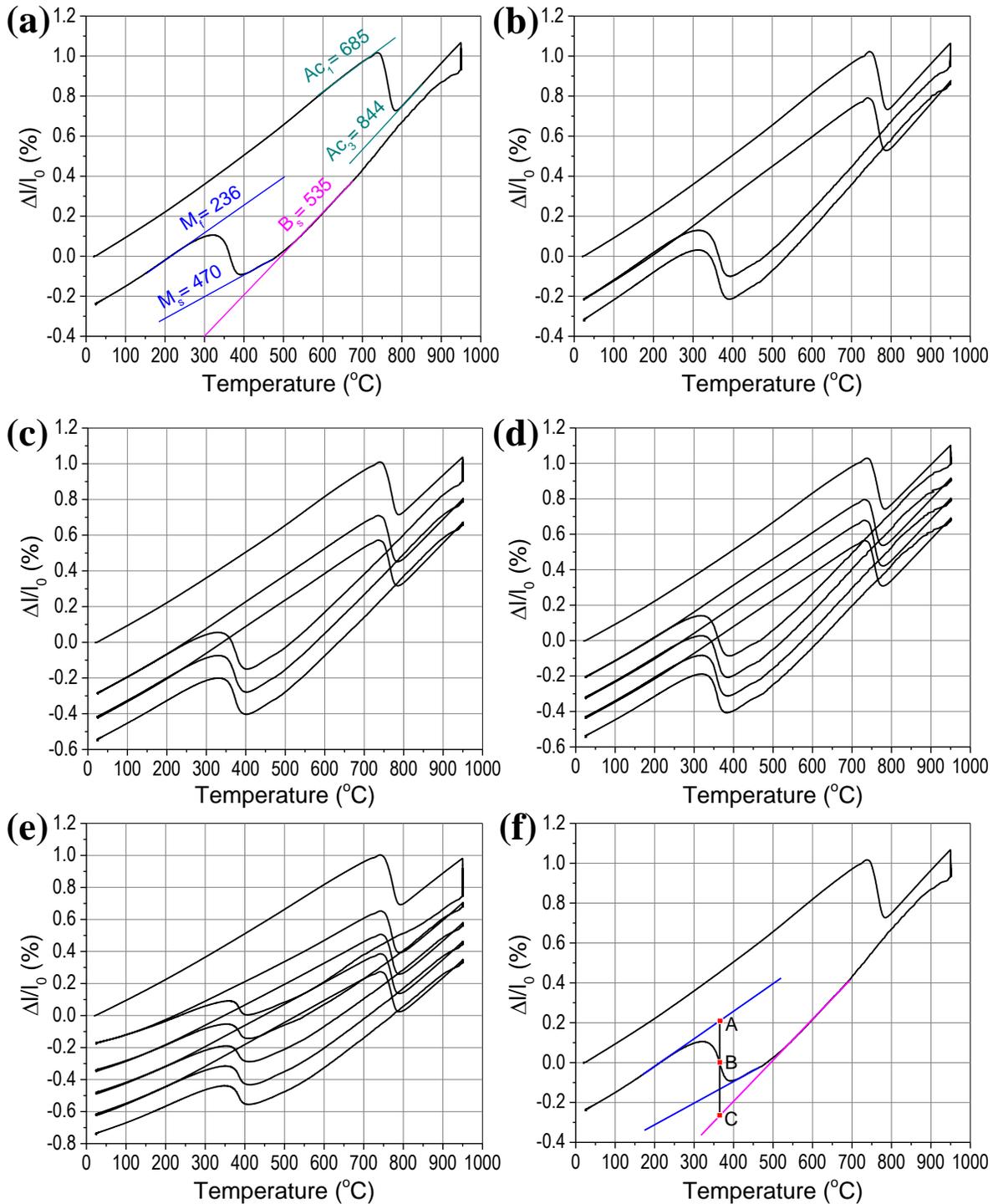

Fig. 2. Dilation as a function temperature measured using quenching dilatometry for (a) 0 cycle (with calculated transformation temperatures indicated), (b) 1 cycle, (c) 2 cycles, (d) 3 cycles, (e) 4 cycles, and (f) illustration of the geometrical method used for the determination of the amount of bainite using the dilatometry curve.

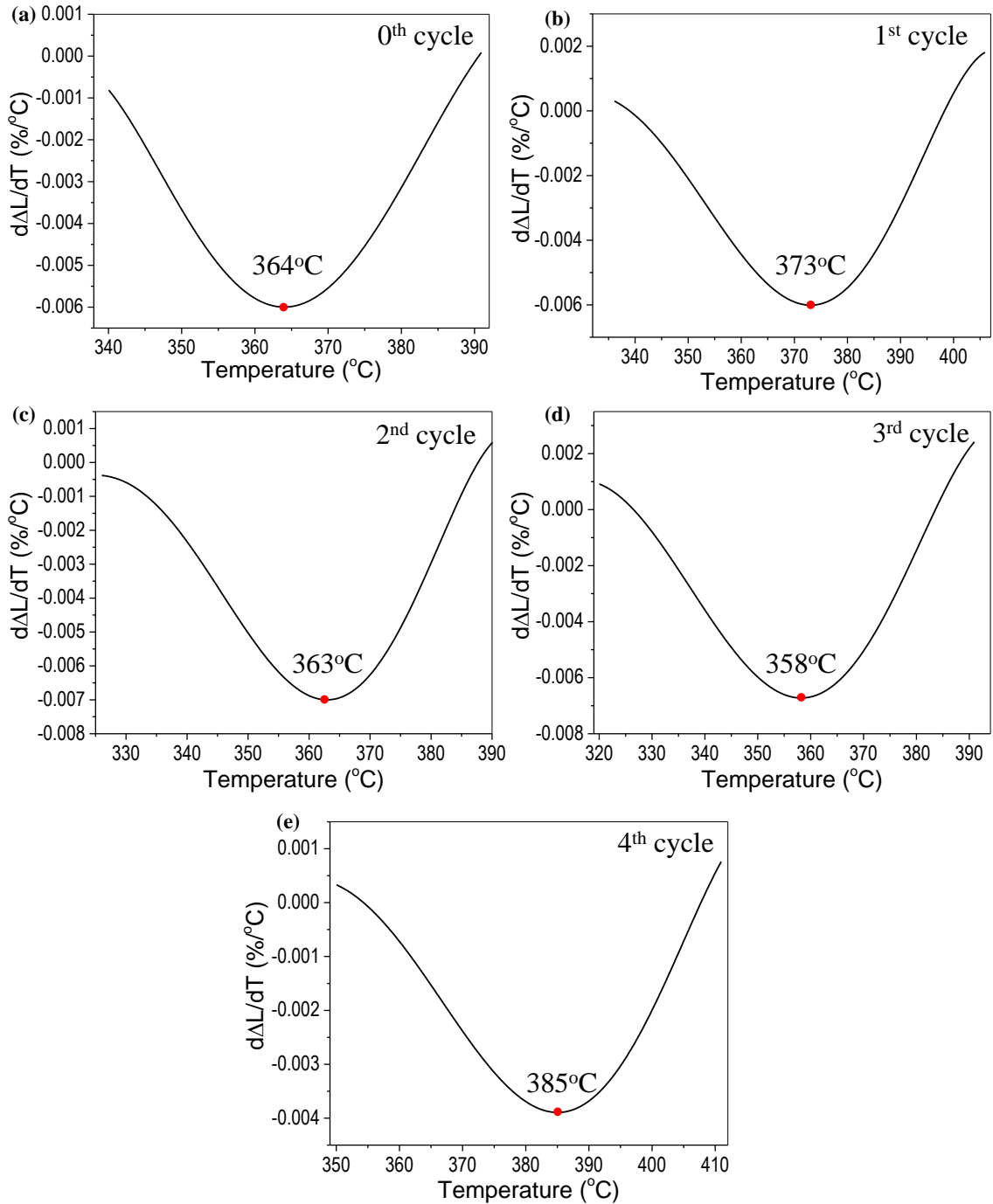

Fig. 3. First derivative of change in length with respect to temperature to identify the inflection point (Point B in Fig. 2(f)) for calculating the bainite fraction from dilatometry curves for (a) 0th cycle, (b) 1st cycle, (c) 2nd cycle, (d) 3rd cycle and (e) 4th cycle of re-austenitization. The inflection point has been indicated in each curve.

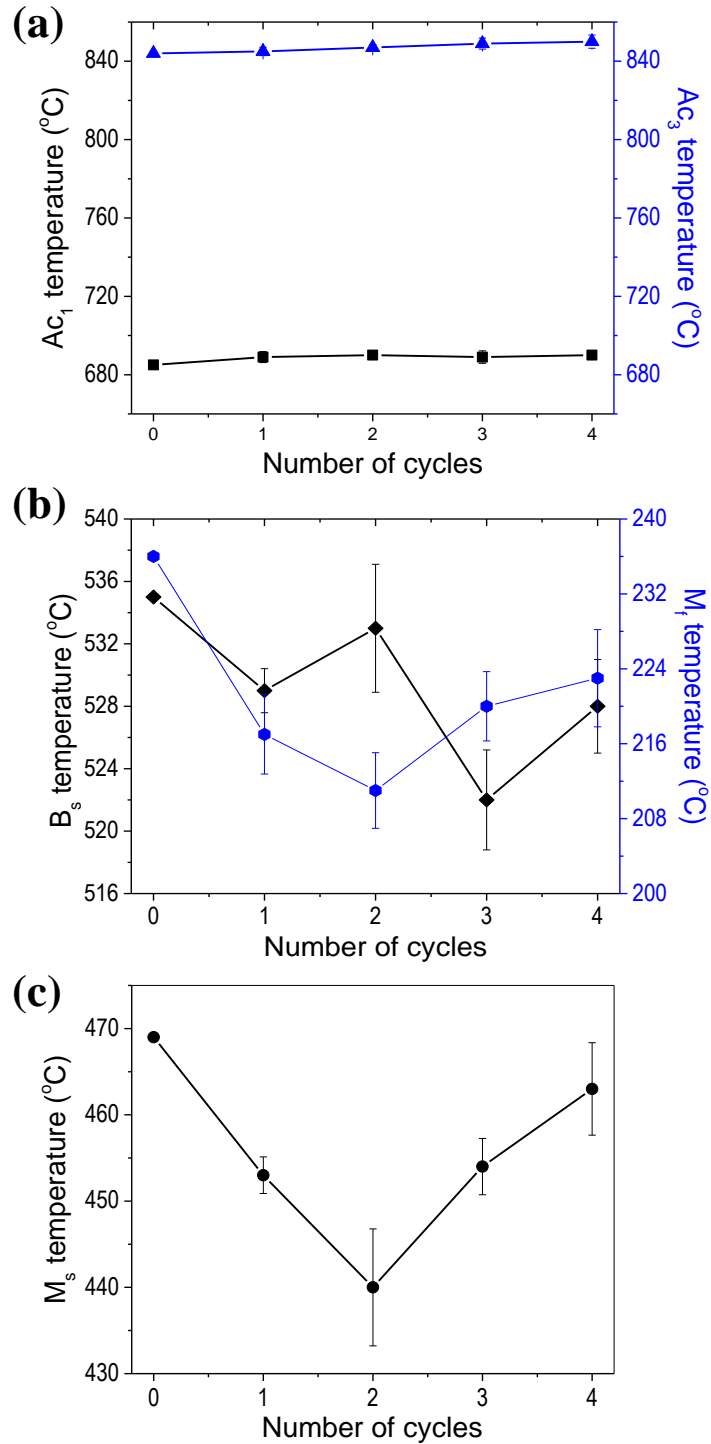

Fig. 4. Transformation temperatures as a function of number of cyclic re-austenitization. (a) $Ac_1$ and $Ac_3$, (b) $B_s$ and $M_f$, and (c) $M_s$ temperatures.

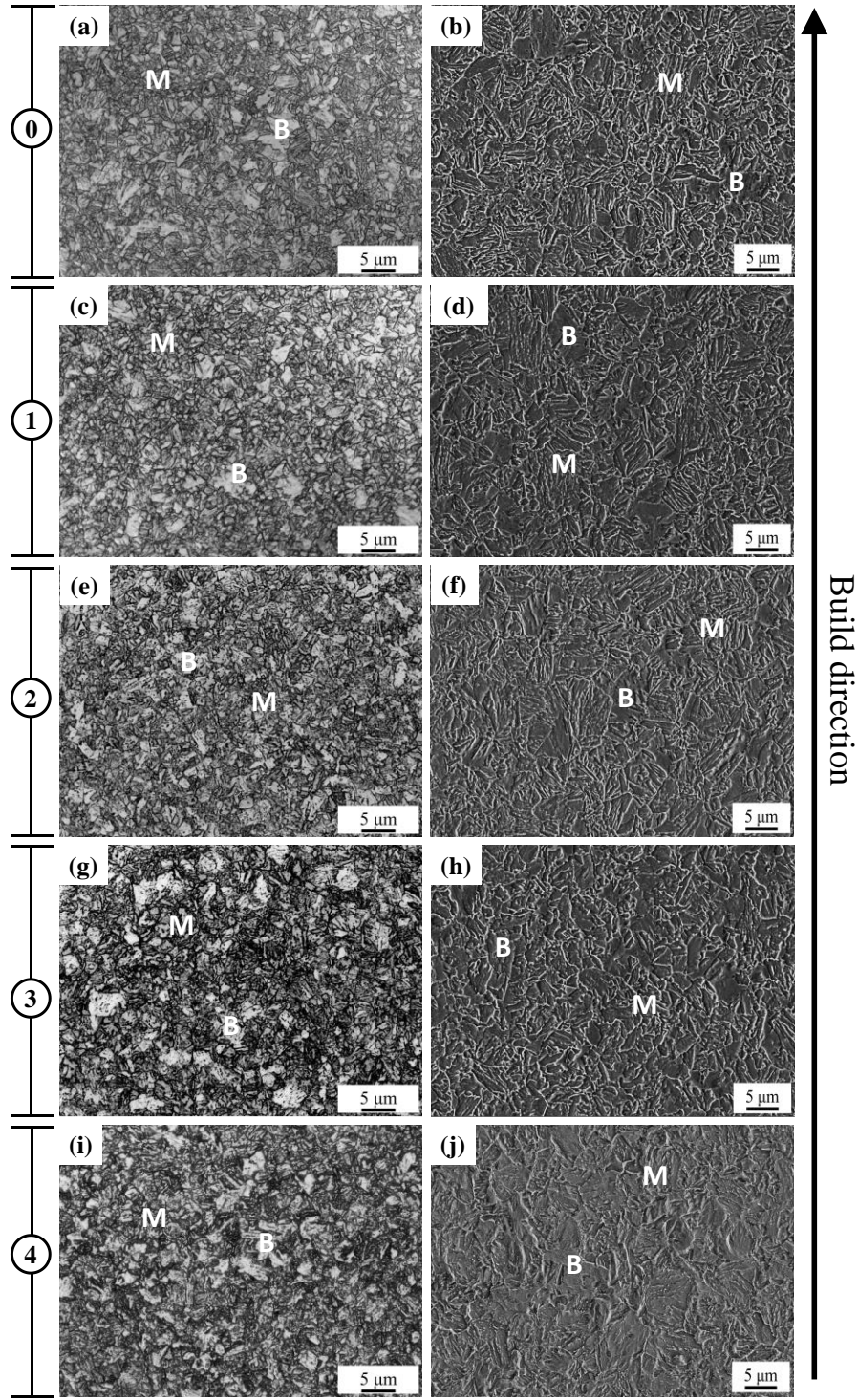

Fig. 5. Optical and SEM micrographs for (a, b) 0 cycle, (c, d) 1 cycle, (e, f) 2 cycles, (g, h) 3 cycles, and (i, j) 4 cycles of cyclic re-austenitization (The index in the left indicates the number of cycles).

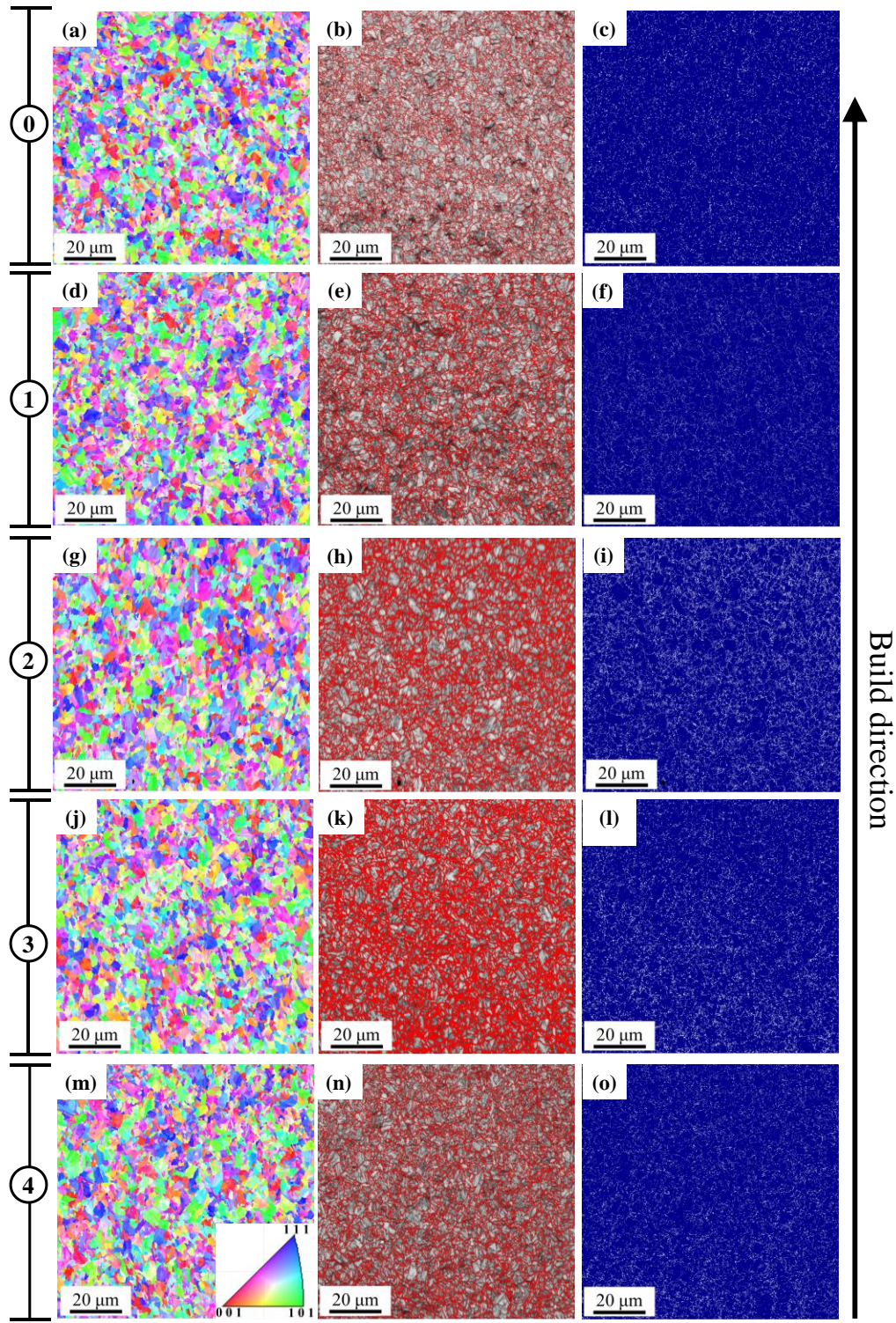

Fig. 6. IPF, IQ and phase maps (blue- BCC and white: FCC) obtained using EBSD for (a-c) 0 cycle, (d-f) 1 cycle, (g-i) 2 cycles, (j-l) 3 cycles, and (m-o) 4 cycles of cyclic re-austenitization (The index in the left indicates the number of cycles).

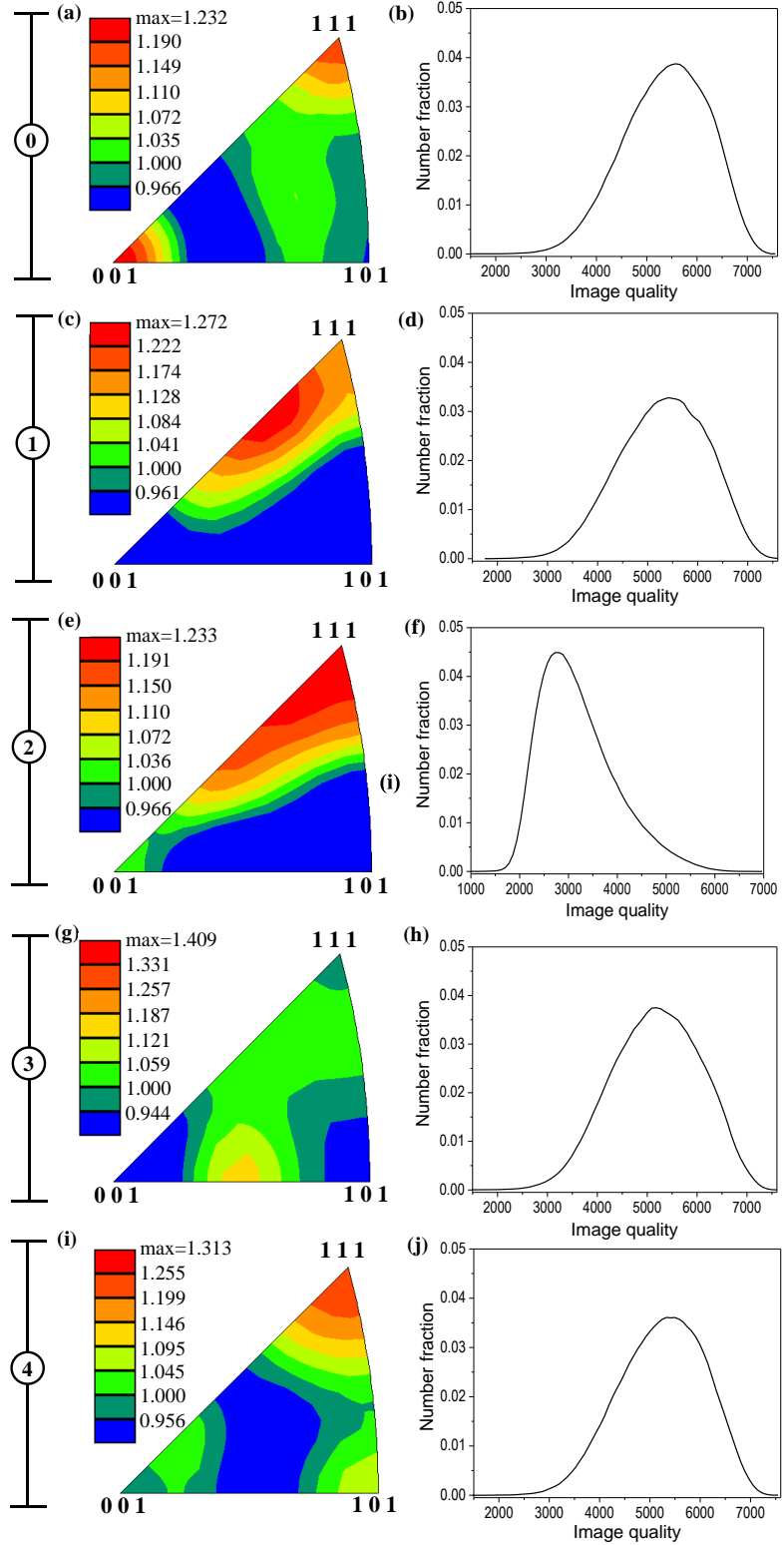

Fig. 7. Texture intensities and IQ plot for BCC phase obtained from EBSD for (a, b) 0 cycle, (c, d) 1 cycle, (e, f) 2 cycles, (g, h) 3 cycles, and (i, j) 4 cycles of re-austenitization (The index in the left indicates the number of cycles).

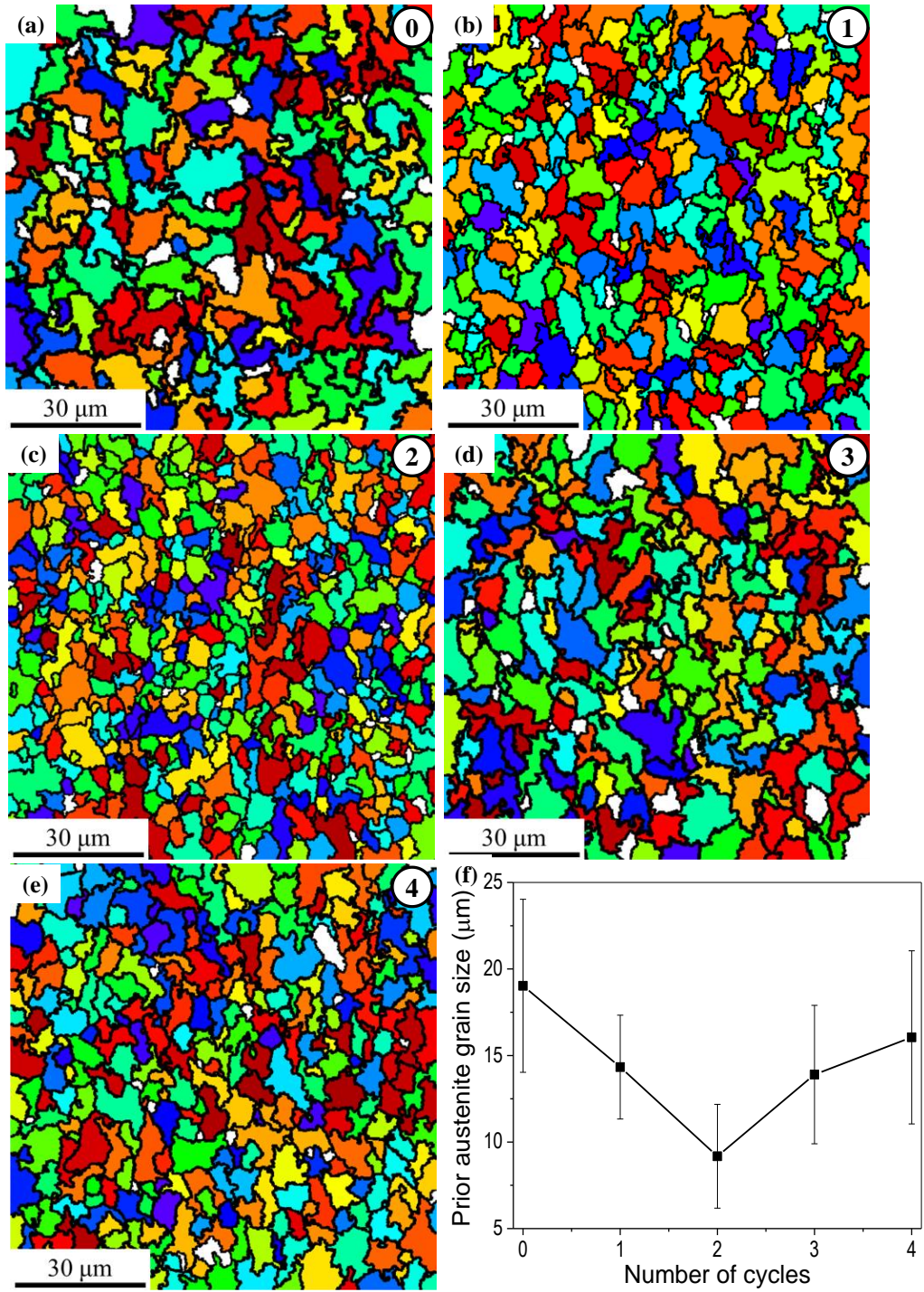

Fig. 8. Reconstructed PAG maps obtained using ARGPE software for (a) 0 cycle, (b) 1 cycle, (c) 2 cycles, (d) 3 cycles, (e) 4 cycles, and (f) PAG size calculated using linear intercept method as a function of number of cycles of re-austenitization (The index in the top right corner indicates the number of cycles).

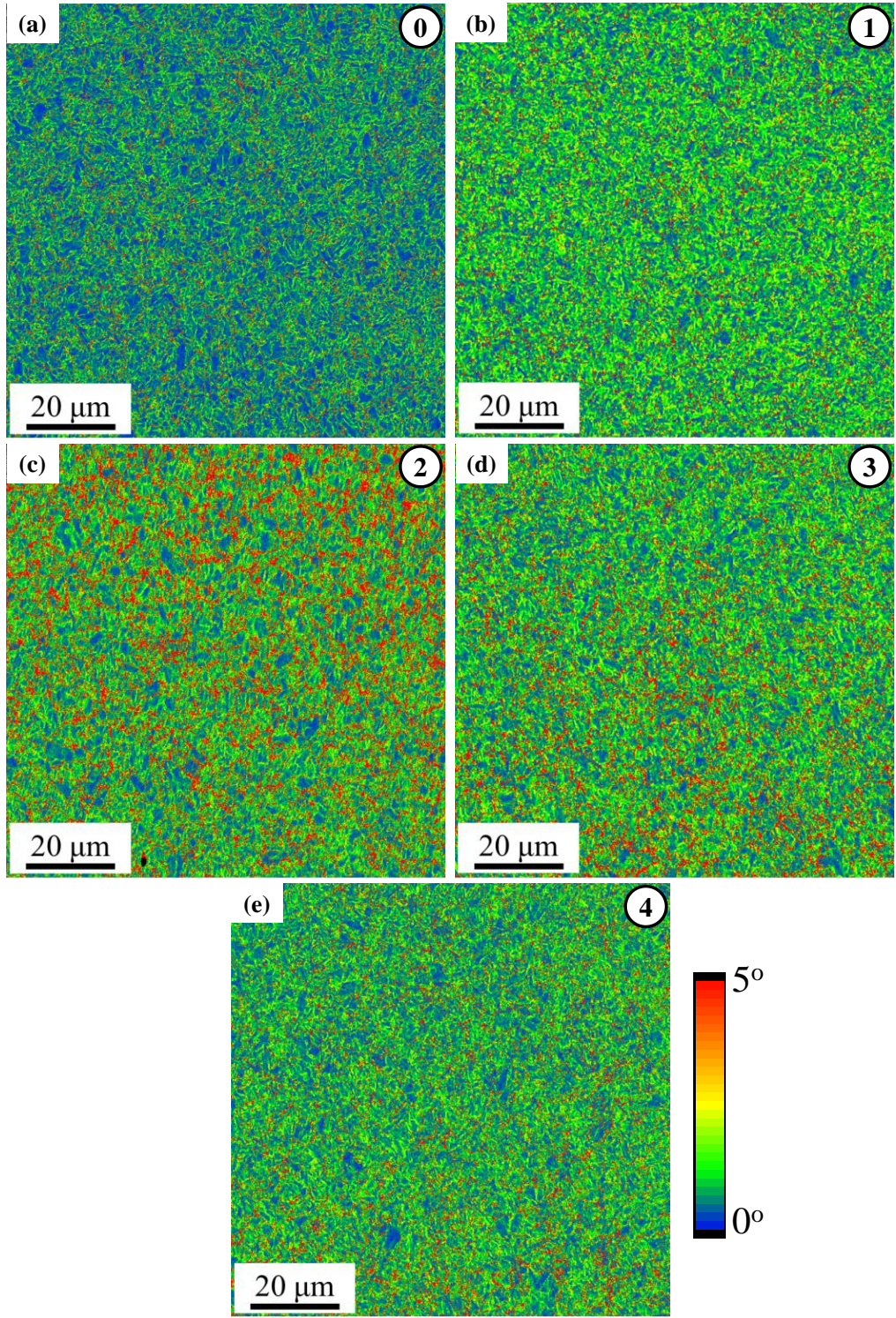

Fig. 9. KAM maps obtained from EBSD for (a) 0 cycle, (b) 1 cycle, (c) 2 cycles, (d) 3 cycles, and (e) 4 cycles of re-austenitization (The index in the top right corner indicates the number of cycles).

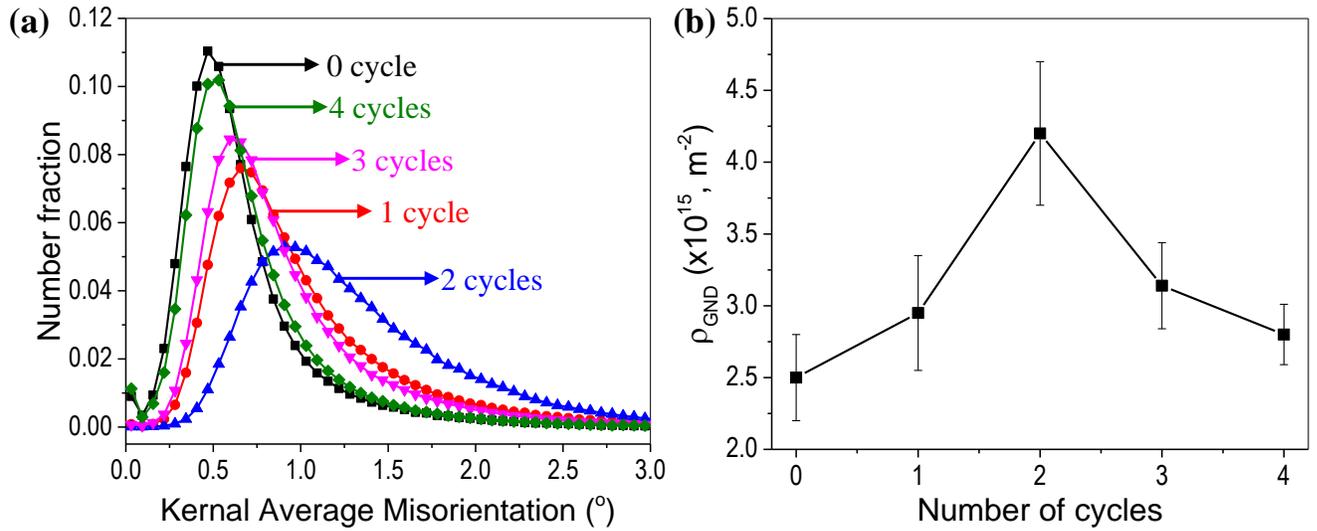

Fig. 10. (a) Average KAM obtained for different cycles of re-austentization and (b) Density of geometrically necessary dislocations as a function of number of cycles of re-austenitization.

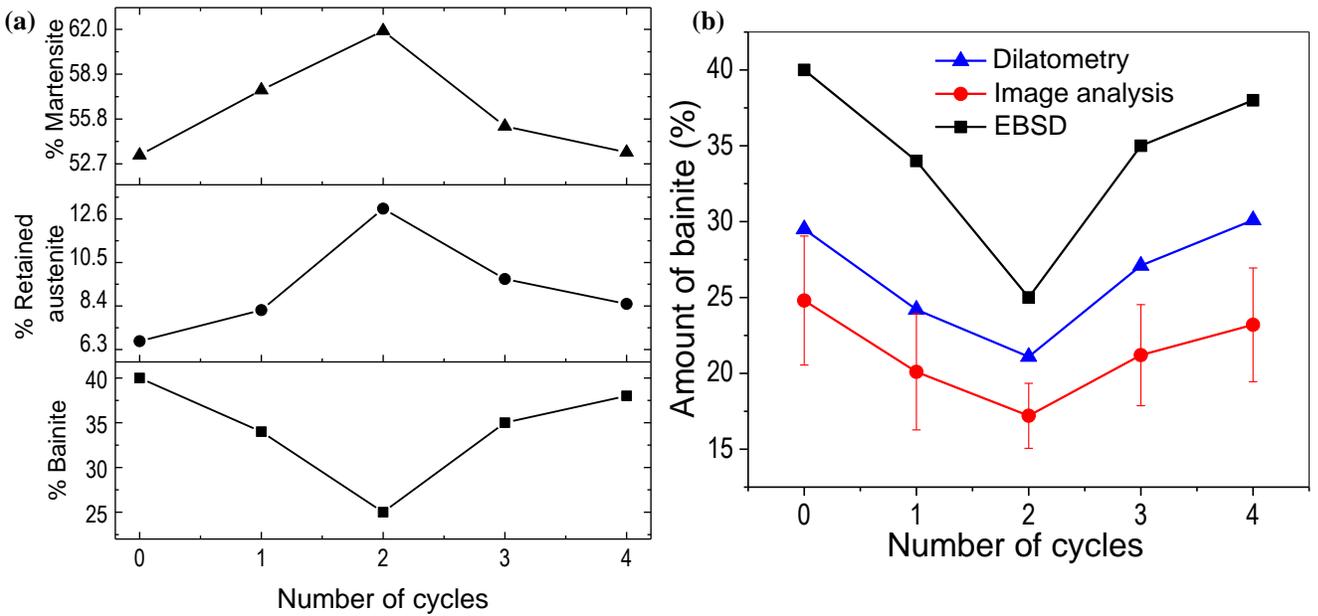

Fig. 11. (a) Amount of phases estimated from EBSD analysis and (b) Amount of bainite calculated using dilatometry, image analysis and EBSD as a function of number of cyclic re-austenitization.

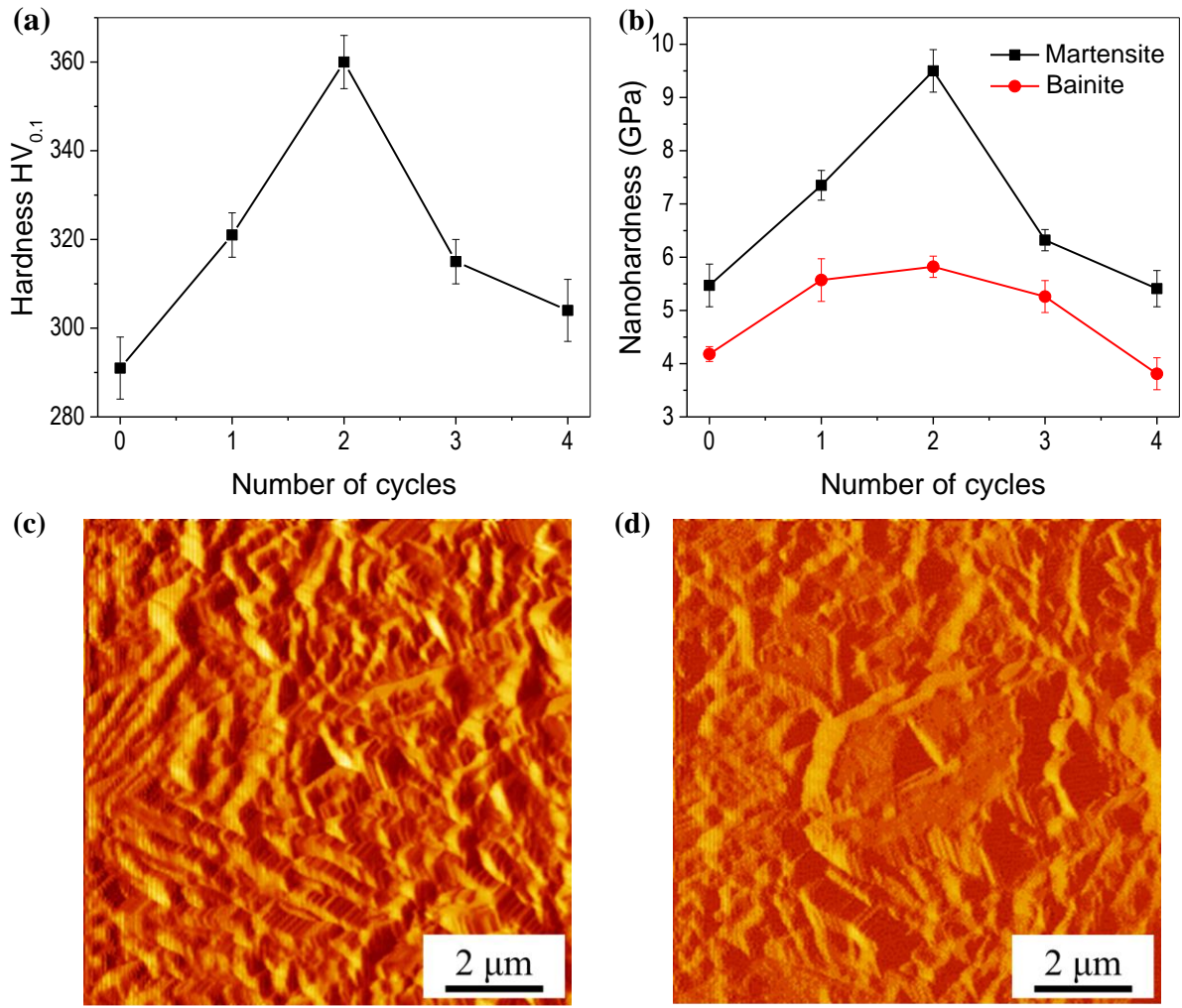

Fig. 12. (a) Microhardness and (b) Nanohardness of the constituent phases as a function of number of cyclic re-austenitization and scanning probe microscopic images of the nanoindents on (c) martensite and (d) bainite.